\documentclass[final,5p,times,twocolumn]{elsarticle}

\usepackage{graphicx}

\usepackage{amssymb}

\journal{Nuclear Instruments and Methods A}

\begin{document}

\begin{frontmatter}



\title{Study of timing performance of Silicon Photomultiplier and application
for a Cherenkov detector}

\author[A,B]{G.S.M.~Ahmed}
\author[A]{P.~B\"uhler}
\author[A]{J.~Marton}
\author[A]{K.~Suzuki}
\address[A]{Stefan Meyer Institute, Austrian Academy of Sciences, Vienna, Austria}
\address[B]{Al-Azhar University, Faculty of Science, Physics Department, Cairo, Egypt}

\begin{abstract}

Silicon photomultipliers are very versatile photo detectors due to their high
photon detection efficiency, fast response, single photon counting capability,
high amplification, and their insensitivity to magnetic fields. At our institute
we are studying the performance of these photo detectors at various operating
conditions. On the basis of the experience in the laboratory we built a
prototype of a timing Cherenkov detector consisting of a quartz radiator with
two $3\times 3$ mm$^2$ MPPCs S10362-33-100C from Hamamatsu Photonics as
photodetectors. The MPPC sensors were operated with Peltier cooling to minimize
thermal noise and to avoid gain drifts. The test measurements at the DA$\Phi$NE
Beam-Test Facility (BTF) at the Laboratori Nazionali di Frascati (LNF) with
pulsed 490 MeV electrons and the results on timing performance with Cherenkov
photons are presented.

\end{abstract}

\begin{keyword}
Silicon photomultiplier, Cherenkov detector, Time resolution
\PACS 29.40.Ka \sep 29.40.Wk

\end{keyword}

\end{frontmatter}


\section{Introduction}

A Silicon Photomultiplier (SiPM), also sometimes referred to as pixelized photon
detector (PPD) is a novel semiconductor photon detector. It consists of a large
matrix of avalanche photodiodes (APD), operated in limited Geiger mode. It has
an intrinsic gain for single photoelectrons of typically $10^6$, which is
comparable to that of a vacuum photomultiplier tube (PMT), and considerably
higher than that of an APD operated in linear
mode~\cite{buzhan_03,sadygov_03,golovin_04}.  Devices from Hamamatsu
Photonics~\cite{hamamatsu} feature typical gain values from $2.5\cdot 10^5$ to
several million, depending on the specific pixel number.  The photo detection
efficiency is peaked at around $440$ nm. Together with its compact size (typical
sensitive area of 1 to 10 mm$^2$)  it makes them ideal for scintillating fiber
readout~\cite{suzuki_09_sipm,stoykov_09}. Insensitivity of the SiPM to a
magnetic field is a big advantage compared to the PMT. The SiPM response, gain,
cross-talk, and noise frequency do not change in the magnetic field within the
measurement accuracy, allowing operation even in high magnetic fields and so a
much more compact design of detectors can be achieved~\cite{buzhan_03,andreev_05}.

SiPMs are also suited for the readout of Cherenkov
detectors~\cite{korpar_08,renker_10}. By combining the fast photon generation
from the Cherenkov process with the high photo-detection efficiency, sensitivity
to low number of photons, and good timing performance of the
device~\cite{suzuki_09_sipm}, such a counter could be used as a beamline
TOF-start counter as needed in high-energy physics experiments. Due to its
compactness and the possibility to operate it in magnetic fields it could be a
valuable alternative to the commonly used scintillating detectors with PMT
readout or CVD Diamond detectors~\cite{CVDD}.

In this paper we present results of an in-beam test of a prototype Cherenkov
detector which uses SiPMs as read-out device. We exposed the detector to an
electron/positron beam (e$^{+/-}$) of $\sim 490$ MeV at the DA$\Phi$NE Beam-Test
Facility (BTF) at the Laboratori Nazionali di Frascati (LNF) and measured the
light outcome as well as the timing resolution of this prototype device.

\section{Prototype Cherenkov detector}

\begin{figure}[hbt]
\centering
\includegraphics[width=0.5\textwidth,keepaspectratio]{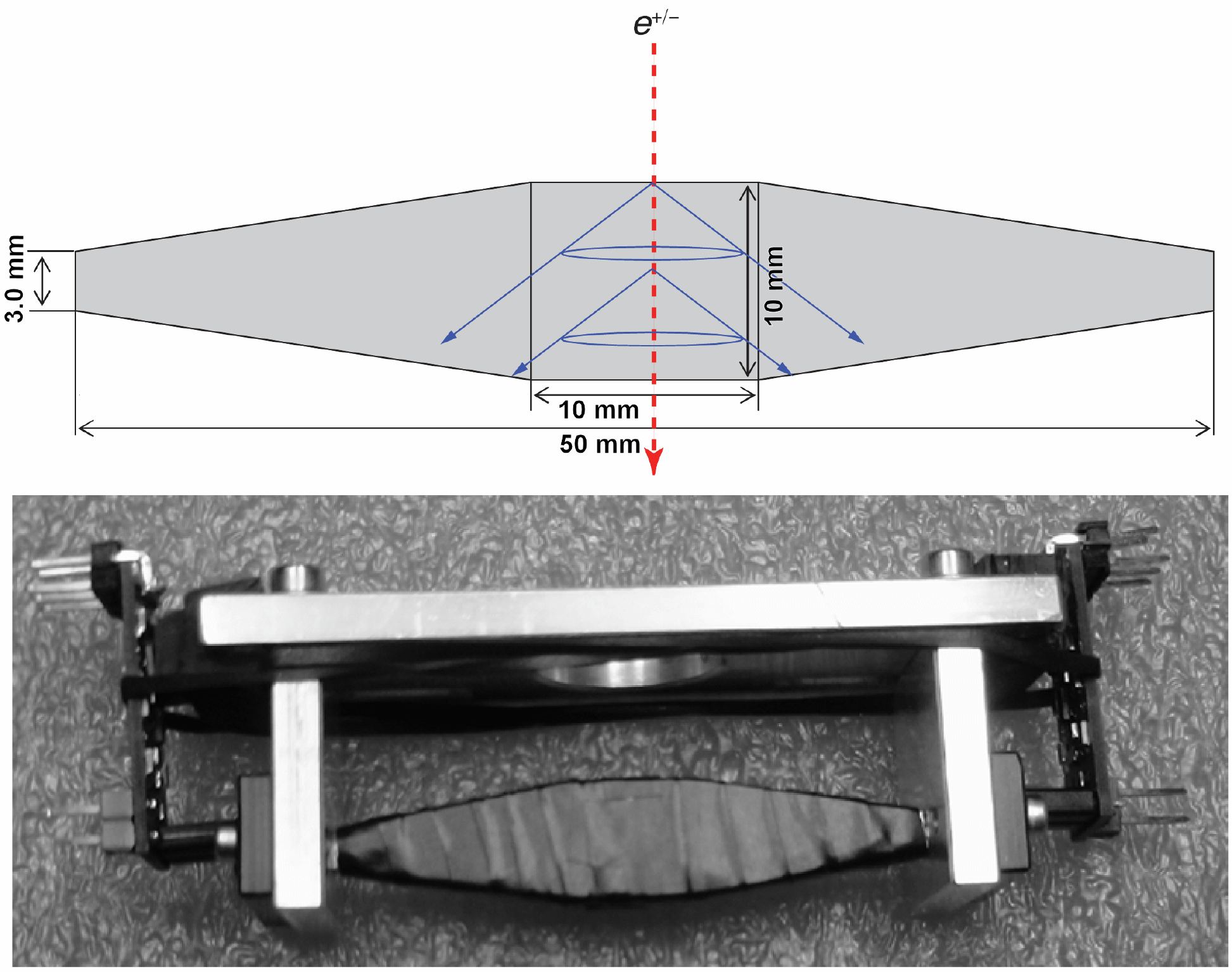}
\caption{Sketch and photograph of the prototype Cherenkov detector. Charged
particles penetrating the radiator (red line in the sketch) produce Cherenkov light
which is guided by total reflection on the internal surfaces of the radiator
(blue line) to its ends. In the photograph the radiator wrapped in black tape,
the aluminum holder, the housings of the SiPMs on both ends of the radiator, and
the preamplifier boards are visible.}
\label{fig1}
\end{figure}

The prototype Cherenkov detector is displayed in figure \ref{fig1}. It consists
of a slab-like radiator of $3$~mm thickness and $50$~mm length. The central part
has a cross section of $10\times 3$~mm$^2$. On both ends it narrows down to a
quadratic surface of $3\times 3$~mm$^2$. The radiator is made of quartz.

The Cherenkov photons produced by a particle traversing the radiator are
supposed to be totally reflected on the internal surfaces of the radiator and to
be guided to its both ends. In order to reduce the losses due to imperfect
reflection the radiator is wrapped with an aluminum foil, and then covered with
light-tight black tape. Both ends of the radiator are optically coupled to a
SiPM which are used to detect the arriving photons. For the presented prototype
detector we used a SiPM from Hamamatsu's series MPPC S10362-33-100C. This
device was primarily selected because of its large area of $3\times 3$ mm$^2$.

This specific SiPM consists of $900$ APDs of $100 \times 100\; \mu$m$^2$ with a
fill factor of $78.5\%$. Its nominal gain is $2.4\cdot 10^6$ and it has a
terminal capacitance of $320$ pF. The maximum photo-detection efficiency is
$\sim 65$\% at 440 nm. Measurements of dark current, dark count rate and timing
performance of this device have been presented elsewhere~\cite{Ahmed_09}.

For testing the Cherenkov detector it is mounted in a light and vacuum tight
aluminum box. The SiPMs are thermally coupled to water-cooled peltier elements,
which allows to regulate the temperature of the device down to approximately
$-20^{\circ}$ C. In order to avoid condensation the box is evacuated to a
pressure of $\approx 10^{-3}$ mbar. The SiPM signals are amplified with fast
preamplifiers, AMP\_0611 from Photonique SA~\cite{photonique}. In order to
minimize electronic noise pick-up the SiPMs are attached directly to the
preamplifier board.

\section{Test at the DA$\Phi$NE Beam-Test Facility (BTF)}

The prototype Cherenkov detector was tested at the electron Beam Test Facility
(BTF) at LNF, Frascati, Italy with e$^{+/-}$ of $\sim 490$ MeV~\cite{BTF}. The
aim was to make a proof of principle and to obtain a first measure of the timing
resolution of this device with a Time-of-Flight (TOF) measurement in a real
accelerator environment.

The BTF is one of the beam lines of the DA$\Phi$NE $\phi$-factory. It is
optimized for the production of electron/positron pulses with a wide range of
multiplicities, including single particles. The maximum repetition rate is $50$
Hz with a maximum particle flux of $1$ kHz, so that at $50$ Hz the maximum
multiplicity is $20$ particles. The beam profile has typical vertical and
horizontal spot size of $d_v = 2$ mm and $d_h = 5$ - $10$ mm.

\begin{figure}[hbt]
\centering
\includegraphics[width=0.5\textwidth,keepaspectratio]{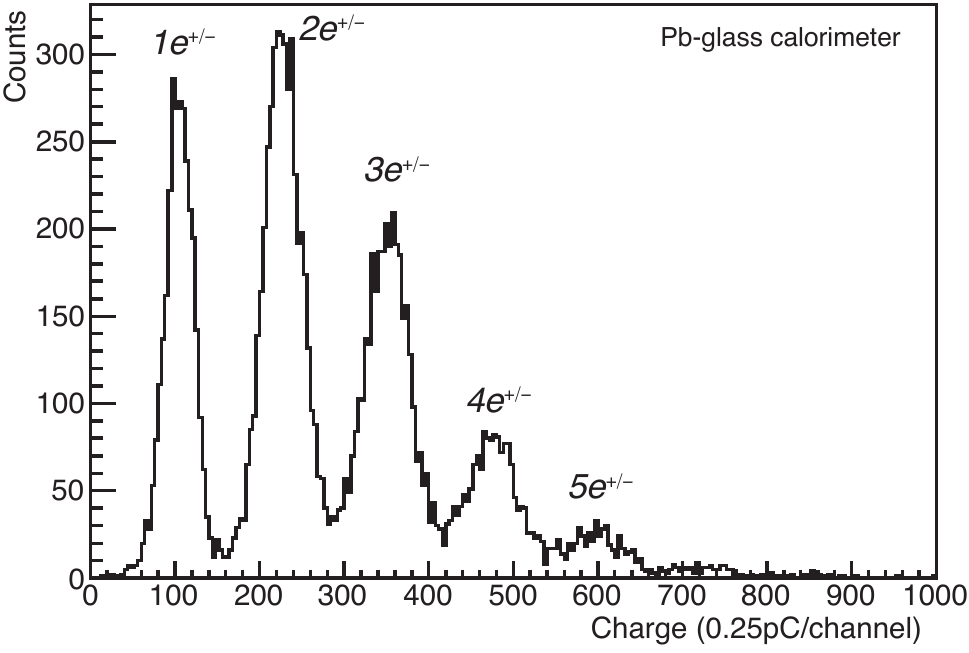}
\caption{Typical energy spectrum measured with the Pb-glass calorimeter. The
separate peaks correspond to beam pulses with different numbers of particles.}
\label{fig2}
\end{figure}

The beam diagnostics elements at BTF include a Pb-glass calorimeter. It is
placed as last detector down stream after the Cherenkov counter in the beam
line. The cross section of the calorimeter is $5\times 5$ cm$^2$ which is
considerably larger than the ones of the other counters used for the test ($\sim
2\times 10$ mm$^2$). The beam particles are totally absorbed in the calorimeter
so that the integrated signal from this detector provides a measure of the beam
particle multiplicity. A typical energy spectrum of the calorimeter is shown in
figure~\ref{fig2}. The separate peaks correspond to beam pulses of $1$, $2$,
$3$ and more beam particles. This information allows to select beam pulses of a
specific multiplicity by software in later data analysis.

\subsection{Experimental setup}

\begin{figure}[hbt]
\centering
\includegraphics[width=0.5\textwidth,keepaspectratio]{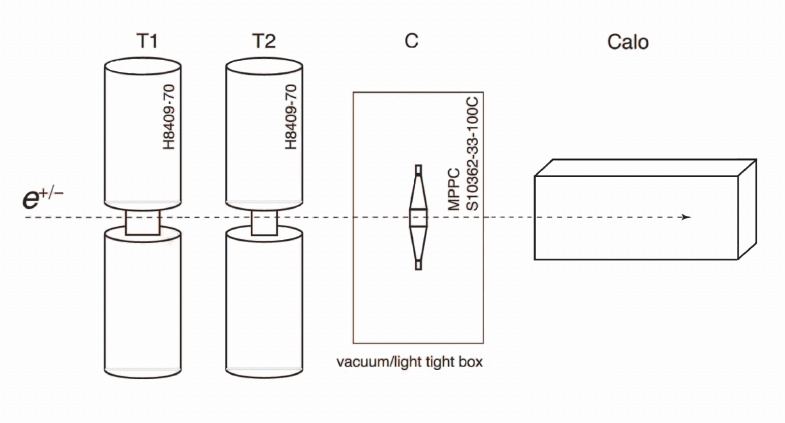}
\caption{Sketch of the experimental setup. The relevant detectors are the
Cherenkov counter C, the reference counters T1 and T2, and the calorimeter
Calo. The reference counters are used to trigger the data taking and to measure
TOF together with the Cherenkov detector. The calorimeter can be used to select
events of a specific beam particle multiplicity.}
\label{fig3}
\end{figure}

Figure~\ref{fig3} shows a schematic drawing of the experimental setup. There
are four detectors relevant for the performed measurements. T1 and T2 are the
reference counters for the TOF measurement, which consist of scintillators of
$20$ and $10$ mm thickness, respectively and are read out on both ends by PMTs
(Hamamatsu H8409-70). The Cherenkov counter C is placed between T2 and the
calorimeter Calo. All detectors were centered on the central beam line. Data
taking was triggered by a coincidence between T1, T2 and a spill gate. The BTF
magnets were set to select e$^{+/-}$ with energies of $\sim 490$ MeV.

\begin{figure}[hbt]
\centering
\centerline{\includegraphics[width=0.3\textwidth,keepaspectratio]{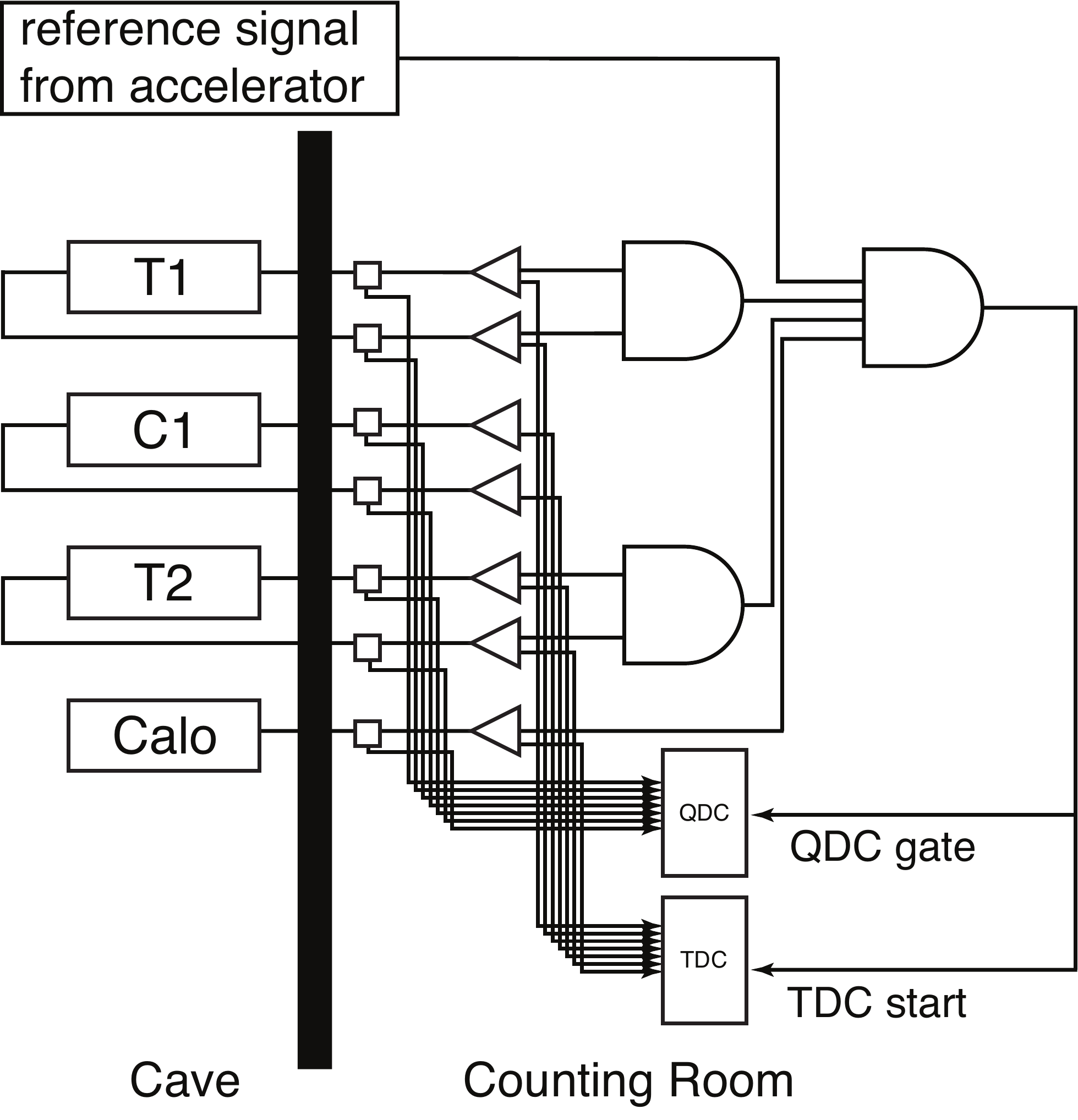}}
\caption{Sketch of the read out electronics. The delay circuit for time
adjustment is not shown here.}
\label{fig4}
\end{figure}

The readout electronics was set up to measure charge spectra and TOF between the
reference counters and the Cherenkov detector. A block scheme of the electronics
setup is shown in figure~\ref{fig4}. The signal output from each counter was
split into two lines. One was connected to a charge-to-digital converter (QDC,
LeCroy ADC-2249W, 0.25pC/channel) for the charge measurement. The other line was
fed into a leading edge discriminator and the outputs of the discriminator were
relayed into a time-to-digital converter (TDC, Phillips 7186 with $25$ ps per
channel resolution) for time measurements and were also used to produce a gate
signal for the QDC and a common-start signal for the TDC. All the QDC and TDC
signals were recorded by a personal computer via Wiener CAMAC-CC32 PCI bus
interface and stored in a hard disk for offline analysis.

\subsection{Measurements}

During the measurements the temperature of the SiPMs was controlled to be
$-16^{\circ}\pm1$ C. The bias voltages of the SiPMs were both set to be $67.2$~V
as an optimization of gain and dark-count, which resulted in a signal pulse
height of $\sim 8$ mV per fired pixel.

\begin{figure}[hbt]
\centering
\includegraphics[width=0.5\textwidth,keepaspectratio]{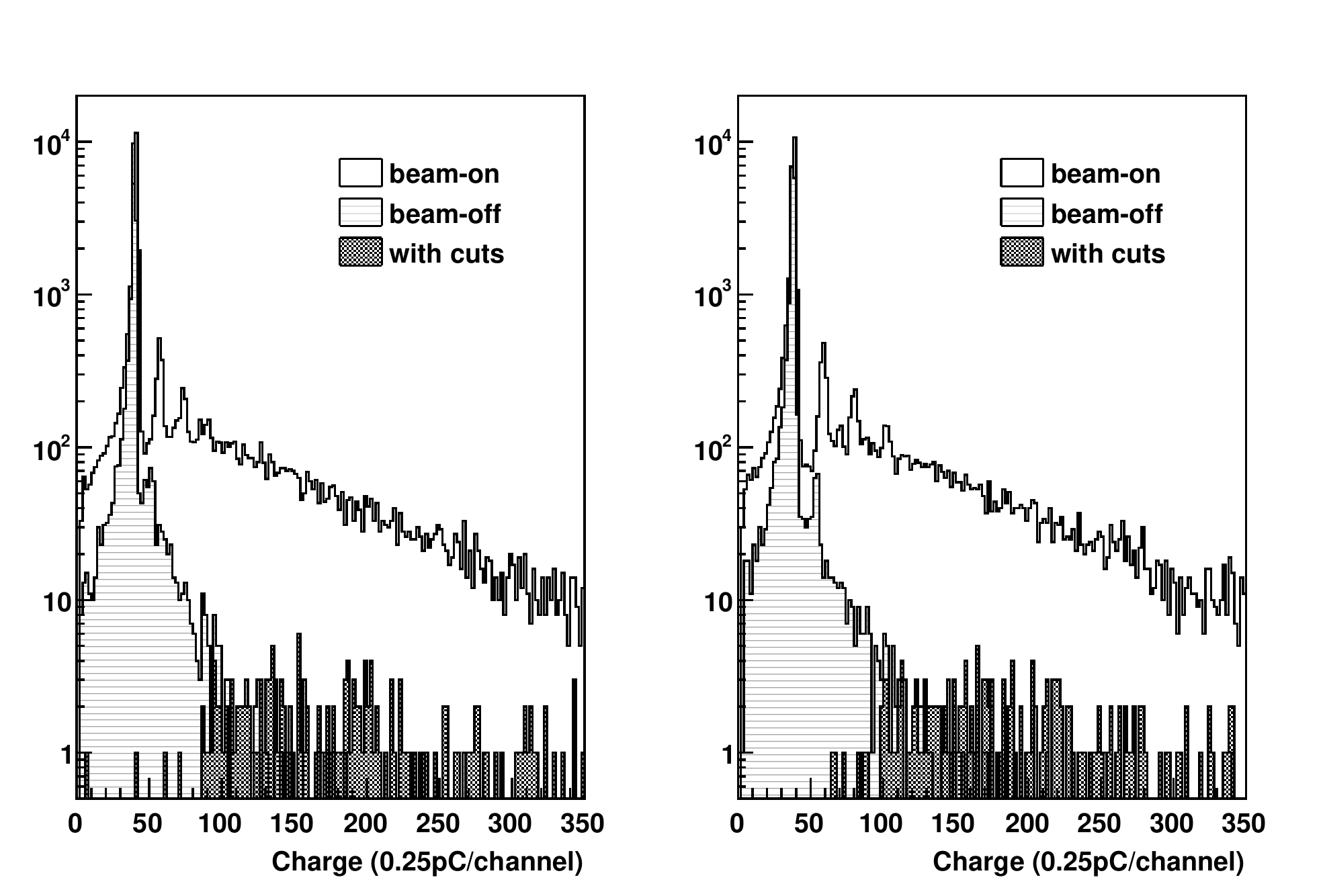}
\caption{Charge distributions of the two SiPMs of the prototype Cherenkov
counter. The empty and striped histograms correspond to beam-on and beam-off
conditions. The filled histogram is remaining signal distribution after
application of all cuts (see text for details).}
\label{fig5}
\end{figure}

Figure \ref{fig5} shows measured charge spectra of both SiPM devices of the
prototype Cherenkov counter. The empty and striped histograms represent the
distributions measured at beam-on (empty) and beam-off (striped) conditions. The
background histogram is scaled so that the intensities in the first peak of both
distributions are equal. Peaks corresponding to $1$, $2$, $3$ and $4$
fired-pixels are visible and well resolved. From the positions of the first four
peaks in each histogram, the photon equivalent number of QDC channels
($QDC_{\mbox{\tiny\it gain}}$) and the QDC offsets ($QDC_{\mbox{\tiny\it
offset}}$) is deduced.  The number of photons ($n_{\mbox{\tiny\it photon}}$)
corresponding to a given QDC channel is given by  $n_{\mbox{\tiny\it photon}} =
(QDC - QDC_{\mbox{\tiny\it offset}})/QDC_{\mbox{\tiny\it gain}}$.

The background distribution is dominated by the one-photon peak and decreases,
in comparison with the beam-on distribution, rapidly towards higher photon
numbers.

The filled histogram represents the signal distribution and is obtained from the
empty histogram after selection of beam pulses with only one particle and events
with good time information in both SiPMs. The number of random coincidences of
larger than two photo-electron equivalent signals is negligible.

The time T(i) of a detector i (with i=T1, T2, C) is defined as the average TDC
value of the corresponding left and right photo-sensors. The TOF between two
detectors i and j is given by TOF(i,j) = T(i)$-$T(j). Using the TOF measurements
of all three possible pairs of detectors allows to disentangle the time
resolution of each single detector and therefore to deduce the timing resolution
of the Cherenkov counter.

For the data analysis, we selected single particle (e$^{+/-}$) events using the
Calo energy spectrum. In addition a cut was applied to select those events in
which both SiPMs of the Cherenkov counter had a proper timing information
registered in the TDC. Note, that the Cherenkov detector did not join the
trigger decision and therefore this cut reduced the number of available events
drastically. Correction for slewing was applied to all counters. For the
selected events the average number of photo-electrons detected in both SiPMs was
estimated to be $\sim 8$ with rather broad distribution (filled histogram in
figure \ref{fig5}).

\begin{figure}[hbt]
\centering
\includegraphics[width=0.5\textwidth,keepaspectratio,angle=0]{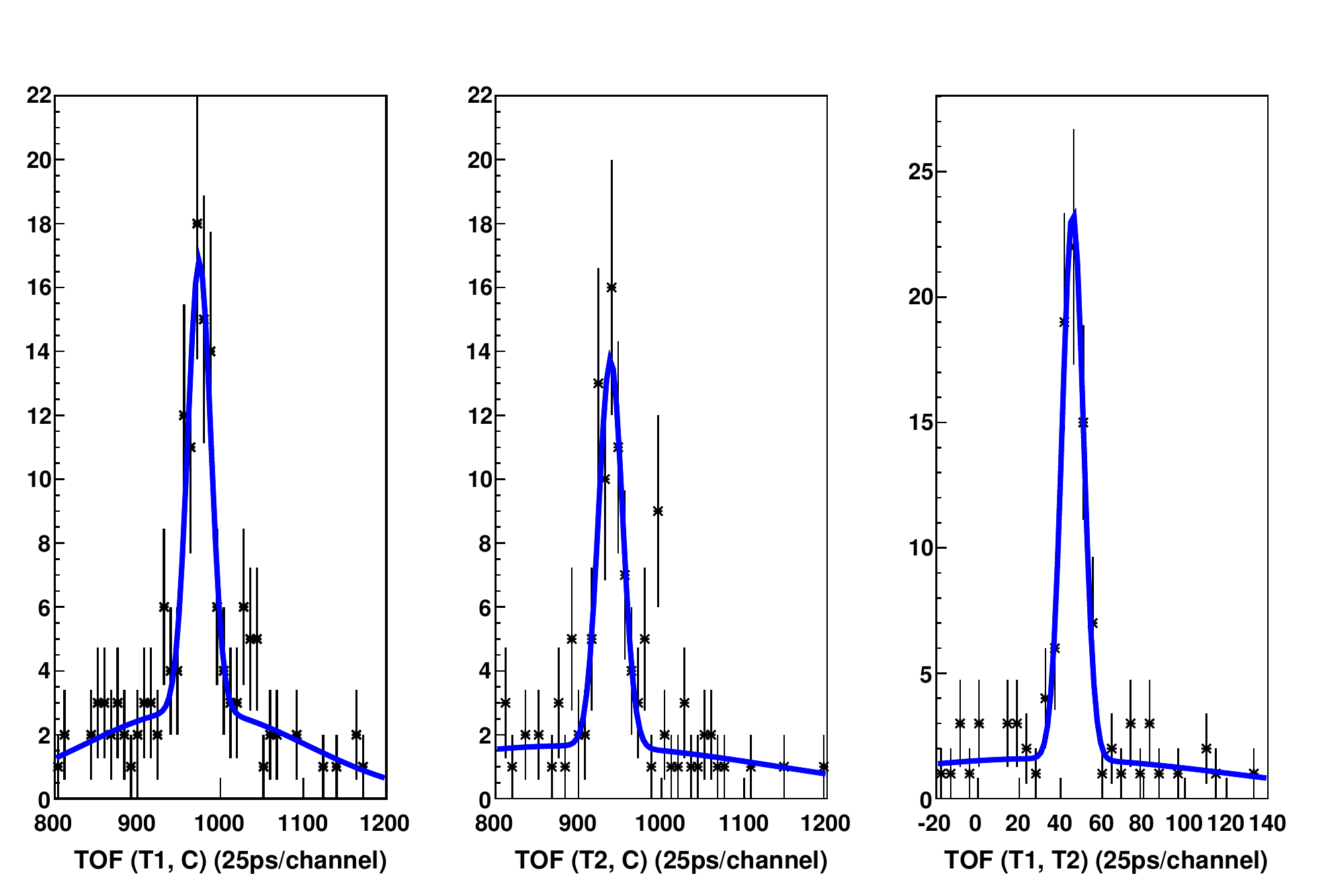}
\caption{Distribution of time difference between pairs of detectors T1 and C
(left panel), T2 and C (middle panel), and T1 and T2 (right panel). The solid
lines are fits to the histograms using two independent gauss functions.}
\label{fig6}
\end{figure}

Figure \ref{fig6} shows the measured TOF distributions between pairs of
detectors T1 and C (left panel), T2 and C (middle panel), and T1 and T2 (right
panel). The solid lines are fits to the histograms using two independent gauss
functions. With the resulting $\sim 360$ ps for TOF(T1, C) and TOF(T2, C) and
$\sim 130$ ps for TOF(T1, T2) a time resolution for the Cherenkov counter of
$\sigma_{T(C)}\sim 350$ ps is obtained. The two reference counters, T1 and T2
have time resolutions of below $100$ ps, which is well below the resolution of
C. The accuracy of this measurement was limited by statistics and we estimate
the error of $\sigma_{T(C)}$ to be in the order of $100$ ps.

\section{Discussion and Conclusions}

We have reported on the beam test of a prototype Cherenkov detector with SiPM
readout. Such a type of detector could be used as a beamline timing counter. In
an experiment with an e$^{+/-}$ beam of $\sim 490$ MeV we detected the Cherenkov
light induced by the particles passing through a slab of quartz with SiPMs and
also measured TOF resolution in combination with reference counters.

The measured time resolution of the detector was found to be rather poor as a
TOF counter ($350\pm 100$ ps). However, it is roughly in agreement with our
previous measurements under laboratory conditions using a short pulse width
laser (Advanced laser diode systems, EIG1000D+PIL040, pulse width 32 ps at wave
length of 408 nm) with the same type of SiPM, pre-amplifiers and readout
electronics at the given number of photons (400 ps at $\sim 8$ photons for one
SiPM, 400/$\sqrt(2) \sim$ 280 ps for the average of two SiPM)~\cite{Ahmed_09}.

The average number of 8 photons detected in each SiPM is according to our
offline measurement in the range of rapidly dropping time resolution as a
function of the number of photons (see e.g.~\cite{Ahmed_09}) and therefore time
resolution is expected to be improved by collecting more photons. Increasing
further the thickness of the radiator from currently 1 cm is not a solution for
all applications and would only help if the generated photons are effectively
guided onto the SiPMs. It must be noted, that the shape of the radiator has not
been optimized for efficient light collection. A simple calculation reveals,
that a 490 MeV electron traveling through 1 cm of quartz generates around 300
photons in the wavelength range between 350 and 800 nm. Assuming an average
photon detection efficiency of the SiPM of 30$\%$ in this wavelength range and
neglecting all further losses, a maximum of 50 photons on each side of the
radiator could be detected. With such a high number of detected photons the time
resolution of the Cherenkov detector would be very much improved.

The other series of SiPM from Hamamatsu, S10362-11 has potentially better
intrinsic timing performance (according to the product catalogue
from~\cite{hamamatsu}), however, due to the smaller sensitive area (1 mm$^2$),
the reduction of photon collection is expected to have a negative impact on the
overall timing performance.

The measured rise time of the amplified signal under the condition of the beam
test is 3 ns. This is significantly slower than the prompt photon generation
from Cherenkov effect and also slower than scintillation light of, for example,
commonly used Bicron BC-408 (time constant 2.1 ns). As the overall timing
performance is a convolution of time response of photon generation, photon
propagation, detector response and readout circuit the overall time resolution
is limited by the slowest component. The above mentioned 3 ns rise time is a
convolution of intrinsic timing response of the SiPM device and the
preamplifiers. We measured that the rise time of the applied preamplifiers,
AMP\_0611 from Photonique SA~\cite{photonique} is 2-3 ns depending on the input
pulse height and therefore contributes significantly to the overall rise time of
the SiPM signal. Thus using a faster preamplifier might also help to improve the
overall time performance.

By improving the design of the radiator and using dedicated electronics for
amplification and readout of the SiPM signals we are optimistic to be able to
improve these results in future.

\section{Acknowledgments}
We wish to express our thanks to Bruno Buonomo, Giovanni Mazzitelli and all
operators of the BTF facility for their help and support. This work is partly
supported by INTAS (project 05-1000008-8114) and Hadronphysics2 (project
227431). One of us (G.A.) acknowledges the support by the Egyptian Ministry of
higher education.





\begin{thebibliography}{00}

\bibitem{buzhan_03}
P. Buzhan {\it et al.}, 
Nucl. Instr.and Methods A 504 (2003) 48.

\bibitem{sadygov_03}
Z. Ya. Sadygov {\it et al.}, 
Nucl. Instr. and Methods A 504 (2003) 301.

\bibitem{golovin_04}
V. Golovin and V. Saveliev,
Nucl. Instr. and Methods A 518 (2004) 560.


\bibitem{hamamatsu}
Hamamatsu Photonics K.K., Japan (http://www.hamamatsu.com/).

\bibitem{korpar_08}
S.~Korpar,
Nucl. Instr. and Methods A 594 (2008) 13.

\bibitem{renker_10}
D. Renker,
J. Inst. 5 (2010) P01001.

\bibitem{suzuki_09_sipm}
K.~Suzuki, P.~B\"uhler, S.~Fossati, J.~Marton, M.~Schafhauser and J.~Zmeskal,
Nucl. Instr. and Methods A 610 (2009) 75.

\bibitem{stoykov_09}
A. Stoykov, R. Scheuermann, T. Prokscha, Ch. Buehler and Z.Ya. Sadygov,
Nucl. Instr. and Methods A (in press)

\bibitem{andreev_05}
V. Andreev {\it et al.}, 
Nucl. Instr. and Methods A 540 (2005) 368

\bibitem{suzuki_09_fopi}
K.~Suzuki {\it et al.}, Nucl. Phys. A 827 (2009) 312.

\bibitem{CVDD}
See for example a recent review in Nuclear Physics News Vol. 19 (2009) No.2 by
E.~Berdermann.

\bibitem{Ahmed_09}
G.S.M. Ahmed, J. Marton, M. Schafhauser, K. Suzuki and P. B\"uhler,
J. Inst. 4 (2009) P09004.


\bibitem{photonique}
Photonique SA, Switzerland (http://www.photonique.ch).



\bibitem{BTF}
G.~Mazzitelli {\it et al.}, Nucl. Instr. and Meth. A 515 (2003) 524,\\
See also the web page http://www.lnf.infn.it/acceleratori/btf/publications.html.

\end{thebibliography}
\end{document}